\documentclass[twocolumn,aps,showpacs,preprintnumbers,amsmath,amssymb,unsortedaddress,superscriptaddress]{revtex4}
\usepackage{graphicx,epsf,epsfig,ulem}
\usepackage{bm}
\usepackage{subfigure}
\usepackage{color}
\usepackage{ulem}
\usepackage{amsmath}

\begin{document}
\title{Voltage controlled spin precession}
\bigskip
\author{A. N. M. Zainuddin}
\affiliation{School of Electrical and Computer Engineering, Purdue
University, West Lafayette, IN 47907, USA}
\email{azainudd@purdue.edu}
\author{S. Hong}
\affiliation{School of Electrical and Computer Engineering, Purdue
University, West Lafayette, IN 47907, USA}
\author{L. Siddiqui}
\affiliation{School of Electrical and Computer Engineering, Purdue
University, West Lafayette, IN 47907, USA}
\author{S. Datta}
\affiliation{School of Electrical and Computer Engineering, Purdue University, West Lafayette, IN 47907, USA}
\medskip
\widetext

\begin{abstract}
A recent experiment reports a non-local spin-signal that shows oscillatory behavior as a function of gate voltage when the contacts are magnetized along the direction of current flow, but not when they are magnetized perpendicular to the current, in agreement with the predictions from a simple theory. In this paper we first present  a straightforward extension of this theory to include the angular spectrum of electrons and the extended injecting and detecting contacts. The results are in good agreement with those from a non-equilibrium Green function (NEGF)- based model with contact parameters adjusted to fit the experimental contact conductances. They also describe certain aspects of the experiment fairly well, but other aspects deserve further investigation.
\end{abstract}

\maketitle

			

\section{Introduction}
Voltage controlled spin precession (see ~\cite{koo_st_09} and references therein), proposed in 1990 ~\cite{datta_das_90}, posed two difficult challenges namely (1) spin-polarized injection into a semiconducting channel, and (2) gate control of the Rashba spin-orbit interaction (RSO) in the channel~\cite{rso_ref}. The latter was demonstrated by Nitta {\it{et al.}} in 1997 using an inverted InGaAs/InAlAs quantum well with a top gate \cite{nitta_sdh_97}. But spin-polarized injection into a semiconductor proved to be a more difficult challenge which has only recently been overcome through the combined efforts of many groups around the world~\cite{spin_inj_ref}. Very recently, Koo {\it{et al.}} ~\cite{koo_st_09} have combined both ingredients, spin-polarized injection and gate-controlled RSO, into a single experimental structure using a high mobility InAs heterostructure with a top gate interposed between the current contacts and the voltage contacts (Fig.~\ref{expt_bench}(a)). The non-local voltage signal~\cite{vanWees_nl_03} showed an oscillatory behavior when the contacts are magnetized along the direction of current flow, but not when they are magnetized perpendicular to the current flow (Fig. \ref{expt_bench}(b)), as expected from the theory presented in ~\cite{datta_das_90}. Furthermore, it was shown in~\cite{koo_st_09} that the oscillation (see Fig.~ \ref{expt_bench}(b)) is described well by the expression
\begin{equation}
V_{exp}=A\cos\left(\begin{array}{c}\frac{2m^*\alpha(V_G)L}{\hbar^2}+\phi\end{array}\right)
\label{vexp}
\end{equation}
where $m^*$ is the effective mass, $\alpha(V_G)$ is the RSO measured independently from the Shubnikov-de Haas (SDH) beating pattern, and A and $\phi$ are fitting parameters. The oscillation period $2m^*\alpha(V_G)L/\hbar^2$ was derived by Datta and Das~\cite{datta_das_90} for electrons with wavevectors that are purely along the direction of current flow (x-) with $k_Y=0$, noting that `in practice we have an angular spectrum of electrons' and the `effect is reduced as $\vec{k}$ turns away from the x-axis.'

In this paper we will first (Section~\ref{simp_mod_sec}) describe a straightforward extension of the theory in ~\cite{datta_das_90} to include the sum over the angular spectrum $k_Y$ of electrons. The results closely follow those obtained earlier in ref.~\cite{zulicke} which are more general since they include both electron and hole systems. We show that the results from this simple model are in good agreement with those in  (Fig.\ref{expt_bench}(d)) from a non-equilibrium Green function (NEGF) based model with contact parameters adjusted to fit the experimental contact conductances (Section~\ref{negf_mod_sec}). We hope that a careful comparison of experiments will help refine model proposed here and establish this effect on a firm footing, so that it can be used both for fundamental studies as well as for various proposed applications such as spin-filtering, magnetic recording and sensing or quantum computing~\cite{wolf_sbandy_dsarma}. 

\section{Simple model}
\label{simp_mod_sec}
We start from an effective mass Hamiltonian for a two-dimensional conductor with RSO interaction of the form ($\vec{\sigma}$: Pauli spin matrices)
\begin{equation}
H=-\frac{\hbar^2}{2m^*}\left(\begin{array}{c}\frac{\partial^2}{\partial
x^2}+\frac{\partial^2}{\partial
y^2}\end{array}\right)+\alpha(\sigma_X k_y-\sigma_Y k_X)
\label{Hamiltonian}
\end{equation} We neglect Dresselhaus spin-orbit (DSO) coupling since this is believed to be small in structures of this type~\cite{dso_rso_ref}. Eq.~\ref{Hamiltonian} leads to the dispersion relation
\begin{equation}
E=\frac{\hbar^2k^2}{2m^*}\pm \alpha k , k =+\sqrt{k_X^2+k_Y^2}
\label{e-k}
\end{equation}
with the upper and lower signs corresponding to eigenspinors of the form $\{1\quad\pm\exp(i\phi)\}^T$, where $\tan\phi\equiv-k_X/k_Y$. Here, $x$ and $y$ are the longitudinal (or transport) and transverse direction respectively following the co-ordinate system used in ~\cite{koo_st_09}, which is different from that used in ~\cite{datta_das_90}. Assuming periodic boundary conditions in the transverse direction, both $E$ and $k_Y$ are conserved in the absence of scattering and the two eigenmodes have different $k_X$'s so as to satisfy Eq. \ref{e-k} with the upper and lower signs respectively. For small $\alpha$ we can write approximately
\begin{equation}
k_{X-}-k_{X+}\approx\frac{2m^*\alpha}{\hbar^2}\frac{k_0}{\sqrt{k_0^2-k_Y^2}}
\label{theta}
\end{equation} with $k_0\equiv\sqrt{2m^*E}/\hbar$.

\begin{figure}[]
\begin{center}
\includegraphics[width=0.3\textwidth]{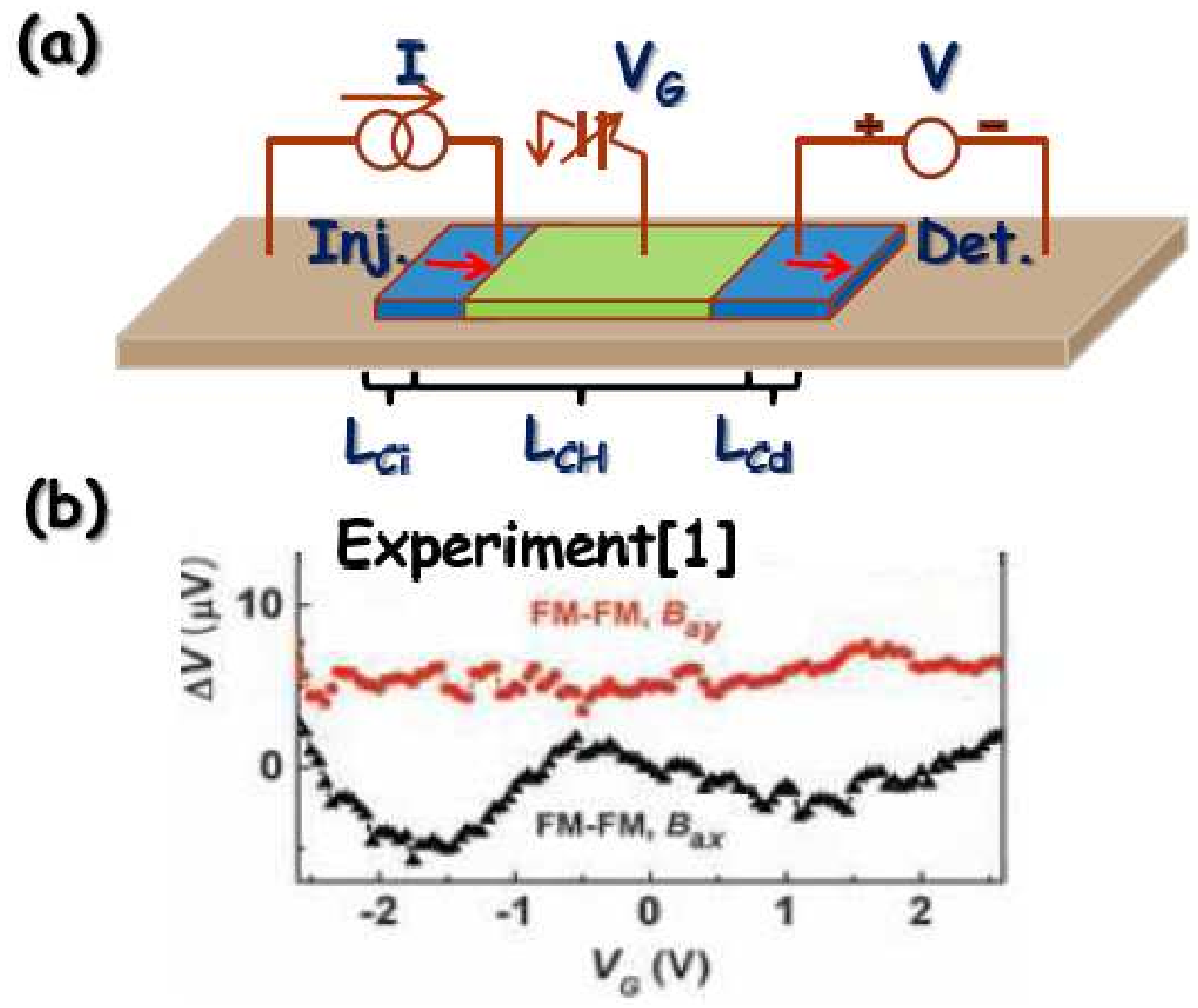}
\includegraphics[width=0.4\textwidth]{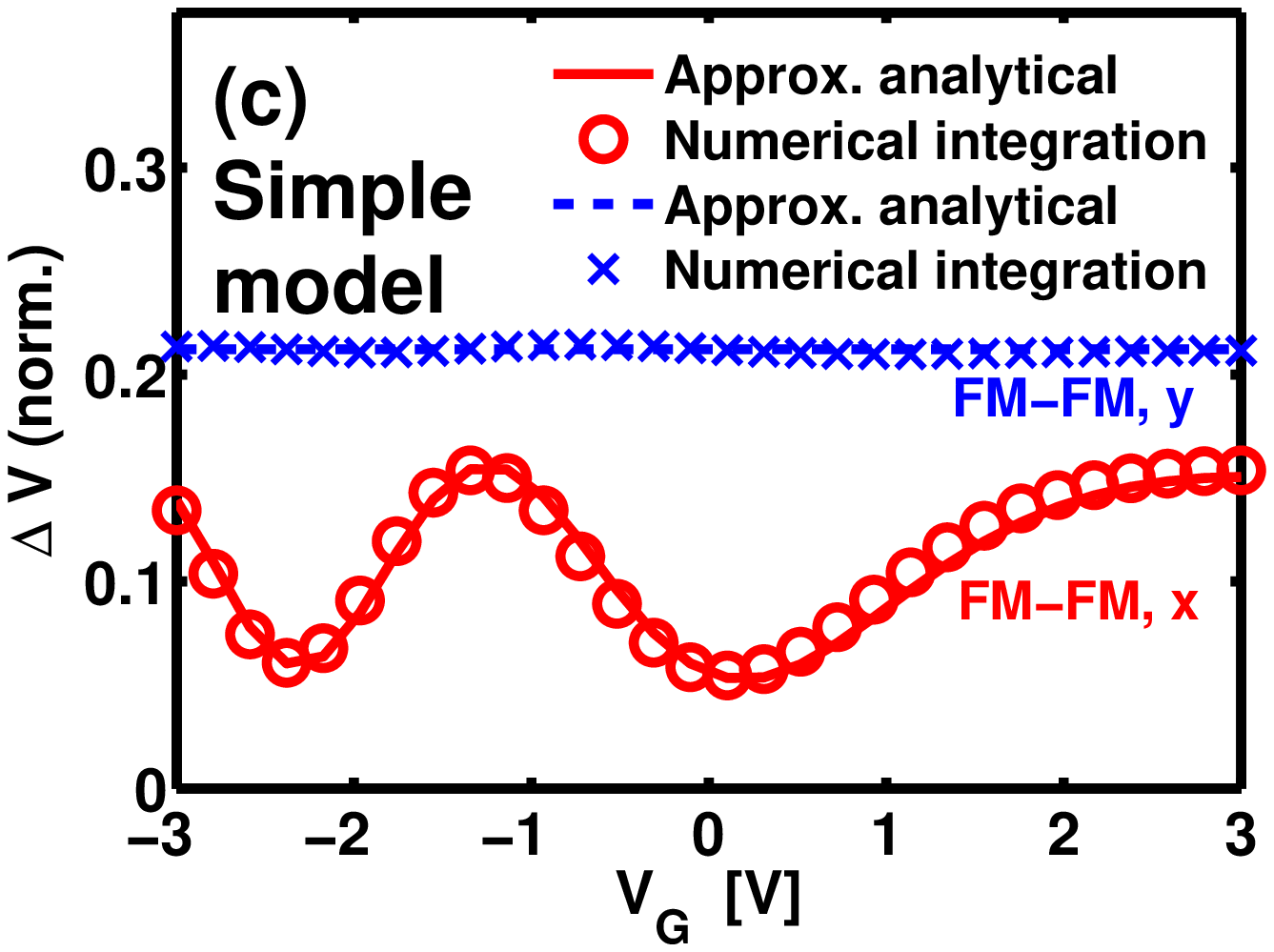}
\includegraphics[width=0.4\textwidth]{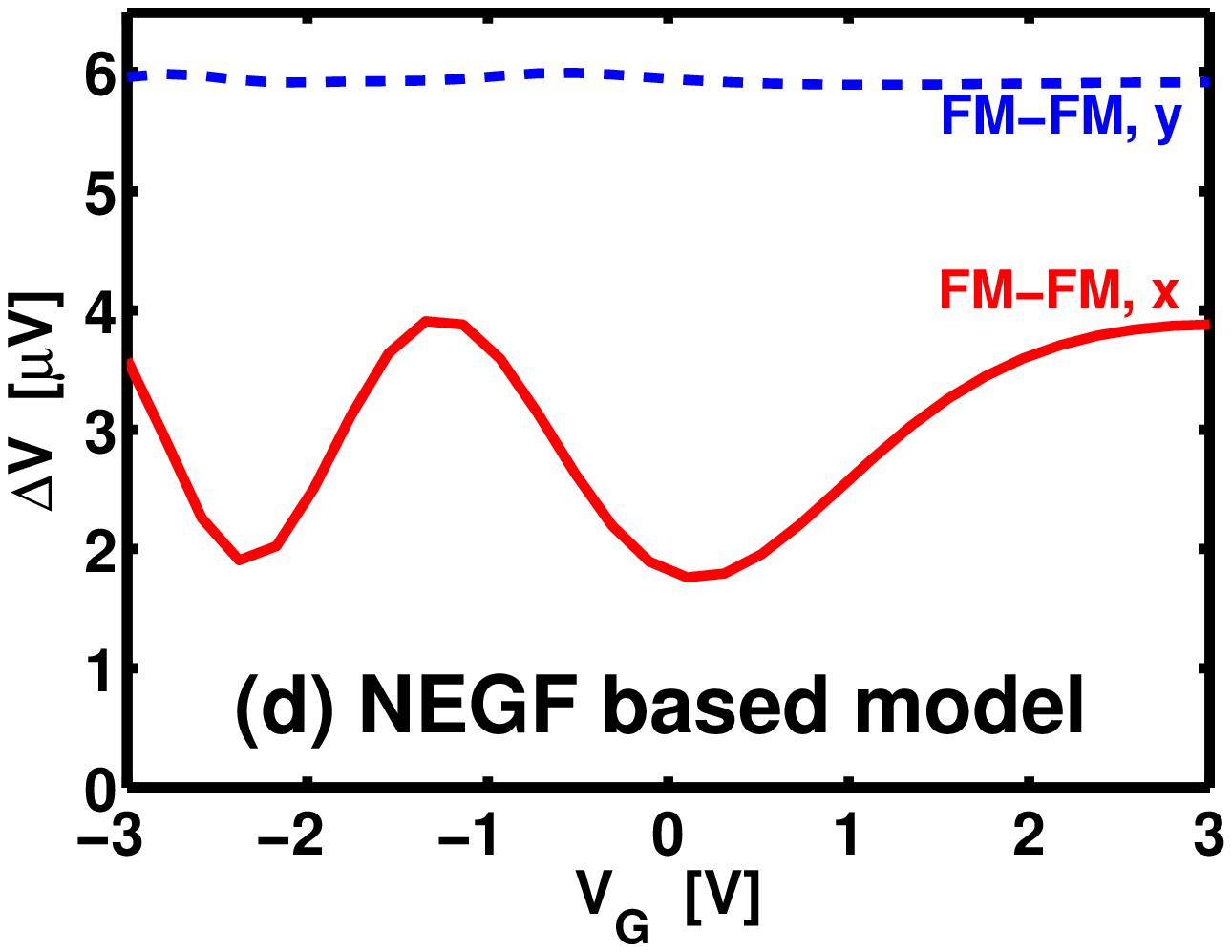}
\end{center}
\caption{(a) Schematic structure and (b) experimentlly observed non-local signal in~\cite{koo_st_09}. Calculated non-local signal for the structure in (a) using: (c) simple analytical model (section~\ref{simp_mod_sec}) and (d) using NEGF based model (section~\ref{negf_mod_sec}). Parameters: $P_C=6.8$\%$=\frac{G_M-G_m}{G_M+G_m}$ and $G_M+G_m = 4\times10^{10}/m^2/\Omega$ ($G_{M(m)}$ being the contact conductance per unit area for majority (minority) spins), carrier density $n_S=2.7\times10^{16}$m$^{-2}$, FM contact lengths $L_{Ci}=0.2\mu$m, $L_{Cd}=0.25\mu$m, FM contact spacing $L_{CH}=1.65\mu$m, width $W=8\mu$m, effective mass $m^*=0.05m_0$.}\label{expt_bench}
\end{figure}

We define the spin-voltage $V_X$ (or $V_Y$) as the difference between the voltages measured in the parallel and the anti-parallel configurations by X- (or Y-) directed injecting and detecting magnets. This is expected to be twice that measured in the parallel configuration using a setup like the one shown in Fig.~\ref{expt_bench}(a) \cite{takahashi} which is exactly how the oscillatory signals $V_X$ in \cite{koo_st_09} are measured, although the spin-valve signal $V_Y$ is measured the same way our $V_Y$ is defined. It is shown in Appendix A (supplementary information) that for a point injector located at $x=0$ and a point detector located at $x=L$, assuming ballistic transport, the voltage signals for X- and Y-directed magnets can be written as ($B$: constant)
\begin{subequations}
\begin{eqnarray}
V_{X0}(E,k_Y)=B\left\{s^2+\left(1-s^2\right)\cos\left(\frac{\theta_L}{\sqrt{1-s^2}}\right)\right\} \label{vx_point} \\
V_{Y0}(E,k_Y)=B\left\{\left(1-s^2\right)+s^2\cos\left(\frac{\theta_L}{\sqrt{1-s^2}}\right)\right\} \label{vy_point}
\end{eqnarray}
\label{v_xy_point}
\end{subequations} where $s\equiv k_Y/k_0=\hbar k_Y/\sqrt{2m^{*}E}$ and $\theta_L=2m^{*}\alpha L/\hbar^2$. The contributions from different $E$, $k_Y$ all act `in parallel' giving a voltage equal to the average. At low temperatures we can average the contributions from all transverse wave-vectors $k_Y$ over the Fermi circle ($E=E_F$) to write
\begin{equation}
V_{X(Y)}=\int^{+k_0}_{-k_0}\frac{dk_y}{2\pi k_0} V_{X0(Y0)}(E_F,k_Y)
\label{vxy_point}
\end{equation} Interestingly, the results obtained from the integration in Eq.~\ref{vxy_point} looks almost exactly like the single cosine result in Eq.~\ref{vexp} that describes the experimental observations. This can be understood by noting that the argument $\theta_L/\sqrt{1-s^2}$ has a stationary point at $s=0$ \cite{stat_point} and we can use the method of stationary phase to write approximately,
\begin{subequations}
\begin{eqnarray}
V_X &\simeq& \frac{B}{3\pi}+\frac{B}{\sqrt{2\pi\theta_L}}\cos(\theta_L+\frac{\pi}{4}) \label{vx_stat_phase} \\
V_Y &\simeq& \frac{2B}{3\pi} \label{vy_stat_phase}
\end{eqnarray}
\label{vxy_stat_phase}
\end{subequations} As shown in Appendix B (supplementary information) these approximations describe the results from the exact integration quite well for $\theta_L\gtrsim 2\pi$ which is true for the range of $\alpha$ and $L$ involved in the experiment. This should help answer some of the concerns raised in a recent comment~\cite{sbandy_com_09}. 

Let us note that the simple results presented above are made possible by our assumption of periodic boundary conditions (PBC) in the y-direction making $k_Y$ a 'good quantum number' like $E$. Most of the prior work, on this topic \cite{hbc_ref}, uses hardwall boundary condition (HBC) which does not seem to permit the simple decoupling of different transverse wavevectors ($k_Y$) due to non-trivial `boundary scattering'. We have checked numerically that the use of HBC does not change the conclusions described above in a significant way although some details are different. Furthermore, one could argue that since the actual boundaries in the experimental structure of \cite{koo_st_09} are relatively far away ($W=8\mu m$) the physics is better captured by a model employing PBC like ours. However, the possible role of boundary scattering deserves further attention.

It is interesting to note the similarities and differences between this simple model for the voltage controlled spinprecession signal and the usual model for the Hanle signal~\cite{hanle_ref}
\begin{equation}
V_H\sim\int^{\infty}_0 dt\frac{e^{-L^2/4Dt}}{\sqrt{4\pi Dt}}\cos\left(\frac{g\mu_B Bt}{\hbar}\right)e^{-t/\tau_S}
\label{vhanle}
\end{equation} The cosine functions in Eqs.~\ref{v_xy_point} can be written as $\cos(2\alpha k_0t/\hbar)$ where $t$ is the transit time $L/v_x=m^*L/\hbar k_0\sqrt{1-s^2}$, showing that $2\alpha k_0$ in our problem plays the role that $g\mu_B B$ plays in the Hanle precession signal. Since we assume ballistic rather than diffusive transport we have a different weighting function for different transit times. But the most important difference is that Hanle signals are typically observed within tens of Gauss around $B=0$ while the experiment we are analyzing has $\alpha$ varying between $8\times 10^{-12}$ and $12\times 10^{-12}$eV-m with $k_0=4.1\times 10^{8}$m$^{-1}$ corresponding to values of $|g|B$ close to $\sim 140$T as noted in~\cite{koo_st_09}. At such high values of $|g|B$, the Hanle signal is usually reduced essentially to zero because of the spread in the transit time `$t$' caused by diffusive transport. One would expect the same in the present case, were it not for ballistic transport. By contrast, the Hanle signal around $B=0$ is relatively robust and it would be interesting to look for an analogous voltage-controlled signal in shorter structures or perhaps in structures where the RSO, $\alpha(V_G)$, can be tuned through $\alpha=0$~\cite{awschalom}.

The simple model here makes no prediction about the amplitude $B$, but it does suggest that the peak-to-peak amplitude of the oscillation in $V_X$ should be $3\pi/\sqrt{2\pi\theta_L}$ times the spin-valve signal $V_Y$. Using $m^{*}=0.05m_0$, $\alpha\simeq 10^{-11}$eV-m, $L=1.65\mu$m, this suggests $V_Y=1.2V_X$(p-p). Experimentally, the p-p oscillatory signal $\sim 6\mu$V which equals $V_X$(p-p)$/2$(since the experiment measures the parallel-antiparallel difference we have defined as $V_X$) but $V_Y$ is only $\sim 6\mu$V. Possible reasons for the discrepancy are discussed at the end of this paper, but here we would like to note that we expect a further reduction in the amplitude of the oscillatory component due to the extension of the injecting and the detecting contacts along $x$ giving rise to a spread in the values of $\theta_L$ in Eqs.~\ref{vx_point} and \ref{vy_point}. We can write
\begin{equation}
\widetilde{V}_x=C_iC_d\frac{B\cos(\theta_0+\theta_i+\theta_d+\pi/4)}{\sqrt{2\pi (\theta_0+\theta_i+\theta_d)}}
\label{vx_contsum}
\end{equation} where $C_i$ and $C_d$ are numbers less than one representing the averaging effects of the injecting and detecting contacts respectively and $\theta_i$, $\theta_d$ are the additional phase-shifts introduced by the injecting and detecting contacts respectively in addition to $\theta_0$. To estimate $\theta_i$, $\theta_d$ or $C_i$, $C_d$ we need to know (1) the spatial uniformity of the injecting and detecting contacts, (2) how the electronic wavefunction evolves under the contacts, and (3) how the RSO $\alpha(V_G)$ varies under the contacts. Regarding point 1 we assume the contacts to be uniform and the NEGF model described next should account for point 2 within this assumption. However, point 3 requires a careful treatment beyond the scope of this paper. Here we simply note that Eq.~\ref{vx_contsum} describes the {\it{shape}} of the oscillatory $V_X(V_g)$ quite well with the following choice: $\theta_0=2m^{*}\alpha(V_G)L_{CH}/\hbar^2$, where $L_{CH}=1.65\mu$m which is the experimental center-to-center distance between contacts and $\theta_{i,d}=m^*\alpha(V_G=0)L_{Ci,d}/\hbar^2$ where $L_{Ci}=0.2\mu$m and $L_{Cd}=0.25\mu$m equal to half the contact widths. The result from Eq.~\ref{vx_contsum} also matches that from the NEGF model (see Fig.~\ref{expt_bench}(d)) to be described next in shape and amplitude if we use $C_{i,d}=\sin(\theta_{i,d})/\theta_{i,d}$ which can be justified if the electronic wavefunction is assumed to remain constant under each contact.

\section{Quantitative NEGF based model}
\label{negf_mod_sec}

\begin{figure}[]
\begin{center}
\includegraphics[width=6.5cm, height=3.0cm]{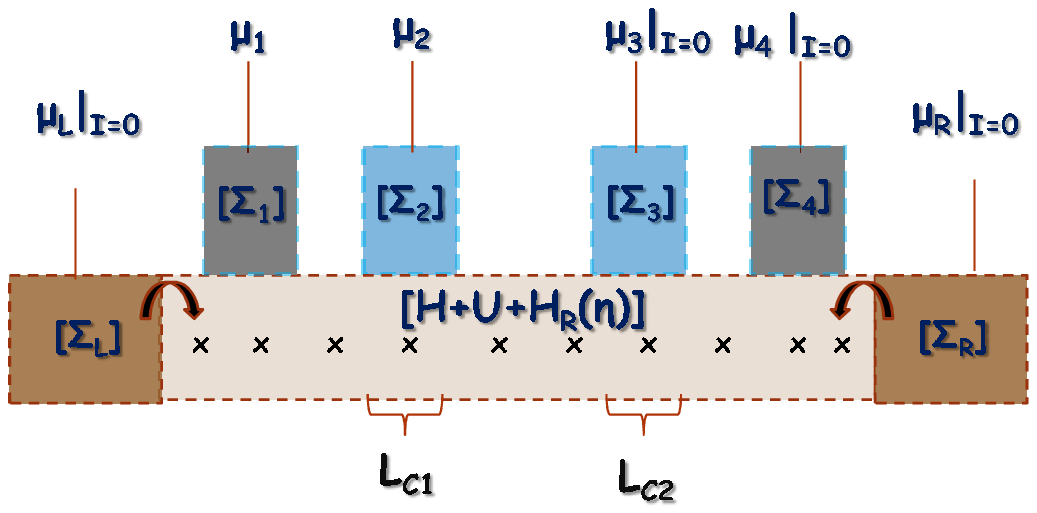}
\end{center}
\caption{NEGF based lateral transport model for the structure in
Fig. \ref{expt_bench}(a), with $\Sigma_2$ and $\Sigma_3$
representing injecting and detecting ferromagnetic contacts,
$\Sigma_1$ and $\Sigma_4$ representing as non-magnetic contacts
(NM), $\Sigma_L$ and $\Sigma_R$ representing the semi-infinite
regions outside the central region}.\label{schematic}
\end{figure}

One way to make the results from the quantum transport model quantitative is to use the NEGF-based method described in detail in \cite{datta3}. The inputs to this model are the Hamiltonian $[H]$ and the self-energy matrices $[\Sigma]$ (Fig~\ref{schematic}). For $H$ we use a discrete version of the one used in Section~\ref{simp_mod_sec} (Eq. \ref{Hamiltonian}), as described in~\cite{datta3} assuming PBC along y as discussed above. We neglect all scattering processes since both the mean free path and the spin-coherence length are believed to be longer than the longitudinal dimensions at low temperatures. To understand the signal decay at higher temperatures will require a consideration of both momentum and spin-relaxation processes, but we leave this for future work. The self-energies for the FM contacts ($\Sigma_2$,$\Sigma_3$) have the form $-(i/2)\gamma[I+P_C\vec{{\sigma}}.\hat{n}]$ where the polarization, $P_C=(G_M-G_m)/(G_M+G_m)$ and $\hat{n}$ is the unit vector in the direction of the magnet. The constant $\gamma=\pi (G_M+G_m)\hbar^3/e^2m^*$ is chosen to give a tunneling conductance equal to the experimental value . The NM contacts ($\Sigma_1$,$\Sigma_4$) are represented similarly with $P_C=0$. Finally, the long extended regions outside the channel at two ends (see Fig. \ref{expt_bench}(a)) are represented by two semi-infinite contacts whose coupling is given by, $\Sigma_{L(R)}=\tau_{L(R)}g_S\tau_{L(R)}^{\dagger}$ where $\tau$ is the spin-dependent coupling matrix between the contact and the channel and $g_S$ is the surface Green's function. The transmission functions are calculated from the NEGF model and contacts 3,4,L and R are treated as voltage probes with zero current (following the approach introduced by Buttiker, see section 9.4, in ~\cite{datta2}). Note that although we are not including scattering processes explicitly, the voltage probes introduce an effective spin-scattering that reduces the signal. Indeed both $V_X$ and $V_Y$ increase significantly if we remove the end regions represented by $\Sigma_L$ and $\Sigma_R$.

For contacts 1 and 2 we adjust the applied potential difference $(\mu_1-\mu_2)$ to obtain a current level equal to the experimental value. The voltage signal $\mu_{3P}$ is obtained from the difference between the $\mu_{3P}$ measured with parallel contacts and $\mu_{3AP}$ measured with anti-parallel contacts. We use a contact conductance of $G_M+G_m=4\times 10^{10}\Omega^{-1}$m$^{-2}$ based on the experimental parameters in~\cite{koo_st_09} and a $P_C=(G_M-G_m)/(G_M+G_m)=0.068$ to match the spin-valve signal, $V_Y$. Fig.~\ref{expt_bench}(d) shows the numerical results obtained using a channel with $\alpha(V_G)$ of length $L_{CH}=1.65\mu$m and contacts with fixed $\alpha(V_G=0)$ of lengths $L_{Ci}=0.2\mu$m and $L_{Cd}=0.25\mu$m. The oscillatory signal matches the experimental observation in shape but the amplitude is smaller. One possibility is that the $P_C$ we use has been calibrated for the spin-valve signals obtained with Y-directed magnets. The same magnets when forced into the X-direction for the oscillatory signals may have a higher effective $P_C$ especially since no anti-parallel measurements are involved. However, to account fully for the discrepancy a significant increase in $P_C$ would be needed and other sources of discrepancy should be investigated.

\section{Discussion}
In summary, we have presented (1) a straightforward extension of the Datta-Das theory ~\cite{datta_das_90} to include the angular spectrum of electrons and the extended contacts, and (2) a more elaborate model that treats the actual non-local experimental structure using a NEGF based approach widely used in nanoelectronics. The simpler theory provides a number of insights and is well-supported by the more elaborate model, identifying several features that deserve further investigation. Specifically it seems that while the experimental oscillation period shows good agreement with theory, the amplitude relative to the spin-valve signal $V_Y$ is larger showing essentially no reduction expected from averaging over the angular spectrum and the extended contacts. Possible reasons for this discrepancy deserve carefull attention. On the theoretical side, it is possible that the contribution from high $k_Y$ components are suppressed because they have shorter effective spin coherence lengths and a purely ballistic theory misses this aspect. Another possibility is that the effective polarization $P_C$ is lower in the antiparallel case which affects only $V_Y$ and not $V_X$. The structure and nature of the injecting and detecting contacts also require careful consederation.

\section{Acknowledgement}
This work is supported by the office of Naval Research under Grant No. N0014-06-1-2005 and the Network for Computational Nanotechnology (NCN).




			


\clearpage
\begin{widetext}
\begin{center}
{\bf{Supplementary information}}
\end{center}
\appendix
\section{Derivation of Eqs. 5a and 5b}
We start from (see Eq. 3)
\begin{eqnarray}
E&=&\frac{\hbar^2}{2m^*}(k_{X+}^2+k_Y^2)+\alpha\sqrt{k_{X+}^2+k_Y^2}
\nonumber \\
&=&\frac{\hbar^2}{2m^*}(k_{X-}^2+k_Y^2)-\alpha\sqrt{k_{X-}^2+k_Y^2}
\label{e-k2}
\end{eqnarray} and
\begin{equation}
\{\psi_+\}\equiv\left\{\begin{array}{c} 1 \\
+\exp(i\phi)\end{array}\right\}, \{\psi_-\}\equiv\left\{\begin{array}{c} 1 \\
-\exp(i\phi)\end{array}\right\} \label{shi}
\end{equation} where $\tan\phi_{+(-)}=-\frac{k_{X+(-)}}{k_Y}$. The incident state $\{\psi_i\}$ is written a linear combination of $\{\psi_{+}\}$ and $\{\psi_{-}\}$
\begin{equation}
\{\psi_i\}=A\{\psi_+\}+B\{\psi_-\}=[\Psi]\left\{\begin{array}{c} A \\
B\end{array}\right\} \label{shi3}
\end{equation} where $[\Psi]\equiv[\{\psi_+\} \{\psi_-\}]$. After propagating from $x=0$ to $x=L$, the final state is written as ($\theta_{+(-)}=k_{X+(-)}L$)
\begin{eqnarray}
\{\psi_f\}&=&A\exp(i\theta_+)\{\psi_+\}+B\exp(i\theta_-)\{\psi_-\} \nonumber \\
&=&[\Psi]\left[\begin{array}{cc} \exp(i\theta_+) &0 \\
0 &\exp(i\theta_-) \end{array}\right]\left\{\begin{array}{c} A \\
B \end{array}\right\} \label{shi4}
\end{eqnarray} Hence we can write, $\{\psi_f\}=[t]\{\psi_i\}$, with
\begin{eqnarray}
[t]=[\Psi]\left[\begin{array}{cc} \exp(i\theta_+) &0 \\
0 &\exp(i\theta_-) \end{array}\right][\Psi]^{-1} \label{t}
\end{eqnarray} where,
\begin{eqnarray}
[\Psi]=\frac{1}{\sqrt{2}}\left[\begin{array}{cc} 1  &1 \\
\exp(i\phi_+) &-\exp(i\phi_-)\end{array}\right] \label{Psi}
\end{eqnarray} Multiplying out the matrices leads to
\begin{eqnarray}
[t]\equiv\frac{\left[\begin{array}{cc} \exp(i\phi_+
+i\theta_-)+\exp(i\phi_- +i\theta_+) &\exp(i\theta_+)
-\exp(i\theta_-) \\
\left\{\exp(i\theta_+) -\exp(i\theta_-)\right\}\exp(i\phi_+ +i\phi_-)
&\exp(i\phi_+ +i\theta_+)+\exp(i\phi_-+i\theta_-)
\end{array}\right]}{\exp(i\phi_+)+\exp(i\phi_-)} \label{t1}
\end{eqnarray} Setting $\phi_+\approx\phi_-\equiv\phi$ (this amounts to ignoring non-orthogonality of "+" and "-" states), the expression simplifies to,
\begin{eqnarray}
[t]\equiv\frac{\left[
\begin{array}{cc}
\exp(i\theta_+)+\exp(i\theta_-) & \left\{\exp(i\theta_+)-\exp(i\theta_-)\right\}\exp(-i\phi) \\
\left\{\exp(i\theta_+) -\exp(i\theta_-)\right\}\exp(i\phi) &\exp(i\theta_+)+\exp(i\theta_-)
\end{array}\right]}{2} \label{t2}
\end{eqnarray} Note that $[t]$ can also be written as
\begin{equation}
[t]=e^{i\left(\theta_++\theta_-\right)/2}\exp\left(i\left[\vec{\sigma}\cdot\hat{n}\right]\Delta\theta/2\right)
\label{tran_rot}
\end{equation} where $\Delta\theta\equiv\theta_+-\theta_-=\theta_L/\sqrt{1-s^2}$ and $\hat{n}$ is a unit vector in the direction of the effective magnetic field: $\hat{n}=\cos\phi\hat{x}+\sin\phi\hat{y}$. This form is intuitively appealing showing the transmission $[t]$ as a product of a simple phase-shift $\exp\left\{i\left(\theta_++\theta_-\right)/2\right\}$ and a rotation around $\hat{n}$ by $\Delta\theta$.

\noindent For z-polarized contacts in parallel configuration,
\begin{eqnarray}
t_{zz}&=&\left\{\begin{array}{cc} 1 &0\end{array}\right\}[t]\left\{\begin{array}{c} 1 \\
0 \end{array}\right\}=t_{11} \nonumber \\
T_{zz}&=&|t_{11}|^2\approx\frac{1+\cos(\theta_+ -\theta_-)}{2}\label{tzz}
\end{eqnarray} and in anti-parallel configuration
\begin{eqnarray}
t_{\bar{z}z}&=&\left\{\begin{array}{cc} 0 &1\end{array}\right\}[t]\left\{\begin{array}{c} 1 \\
0 \end{array}\right\}=t_{21} \nonumber \\
T_{\bar{z}z}&=&|t_{21}|^2\approx\frac{1-\cos(\theta_+
-\theta_-)}{2}\label{tzzz}
\end{eqnarray} For x-polarized contacts in parallel configuration,
\begin{eqnarray}
t_{xx}&=&\frac{1}{2}\left\{\begin{array}{cc} 1 &1\end{array}\right\}[t]\left\{\begin{array}{c} 1 \\
1 \end{array}\right\}=\frac{t_{11}+t_{22}+t_{12}+t_{21}}{2} \nonumber \\
T_{xx}&\thicksim&\left|\frac{(1+\cos\phi)\exp(i\theta_+)+(1-\cos\phi)\exp(\theta_-)}{2}\right|^2 \nonumber \\
&\thicksim&\frac{(1+\cos^2\phi)+\sin^2\phi\cos(\theta_+-\theta_-)}{2}
\label{txx}
\end{eqnarray} and in anti-parallel configuration.
\begin{eqnarray}
t_{\bar{x}x}&=&\frac{1}{2}\left\{\begin{array}{cc} 1 &-1\end{array}\right\}[t]\left\{\begin{array}{c} 1 \\
1 \end{array}\right\}=\frac{t_{11}-t_{22}+t_{12}-t_{21}}{2} \nonumber \\
T_{\bar{x}x}&\thicksim&\frac{(1-\cos^2\phi)-\sin^2\phi\cos(\theta_+-\theta_-)}{2}
\label{txxx}
\end{eqnarray} For y-polarized contacts in parallel configuration,
\begin{eqnarray}
t_{yy}&=&\frac{1}{2}\left\{\begin{array}{cc} 1 &-i\end{array}\right\}[t]\left\{\begin{array}{c} 1 \\
+i \end{array}\right\}=\frac{t_{11}+t_{22}+i(t_{12}-t_{21})}{2} \nonumber \\
T_{yy}&\thicksim&\left|\frac{(1+\sin\phi)\exp(i\theta_+)+(1-\sin\phi)\exp(i\theta_-)}{2}\right|^2 \nonumber \\
&\thicksim&\frac{(1+\sin^2\phi)+\cos^2\phi\cos(\theta_+-\theta_-)}{2}
\label{tyy}
\end{eqnarray} and in anti-parallel configuration.
\begin{eqnarray}
t_{\bar{y}y}&=&\frac{1}{2}\left\{\begin{array}{cc} 1 &+i\end{array}\right\}[t]\left\{\begin{array}{c} 1 \\
+i \end{array}\right\}=\frac{t_{11}-t_{22}+i(t_{12}+t_{21})}{2} \nonumber \\
T_{\bar{y}y}&\thicksim&\frac{(1-\sin^2\phi)-\cos^2\phi\cos(\theta_+-\theta_-)}{2}
\label{tyyy}
\end{eqnarray}

Noting that $\tan\phi\approx-k_X/k_Y$ and $k_0^2\approx
k_X^2+k_Y^2$ we can write,
\begin{eqnarray}
V_Z\sim T_{zz}-T_{\bar{z}z} &=& B\cos\left(\frac{2m^*\alpha L}{\hbar^2}\frac{k_0}{\sqrt{k_0^2-k_Y^2}}\right)\nonumber \\
V_X\sim T_{xx}-T_{\bar{x}x} &=& B\left\{\frac{k_Y^2}{k_0^2}+\left(1 - \frac{k_Y^2}{k_0^2}\right)\cos\left(\frac{2m^*\alpha L}{\hbar^2}\frac{k_0}{\sqrt{k_0^2-k_Y^2}}\right)\right\} \nonumber \\
V_Y\sim T_{yy}-T_{\bar{y}y} &=& B\left\{\left(1 - \frac{k_Y^2}{k_0^2}\right)+\frac{k_Y^2}{k_0^2}\cos\left(\frac{2m^*\alpha L}{\hbar^2}\frac{k_0}{\sqrt{k_0^2-k_Y^2}}\right)\right\} \nonumber \\
\label{spin_sig_T}
\end{eqnarray}

\clearpage
\section{Derivation of Eqs. 7a and 7b}

\begin{figure}[h!]
\begin{center}
\includegraphics[width=0.45\textwidth]{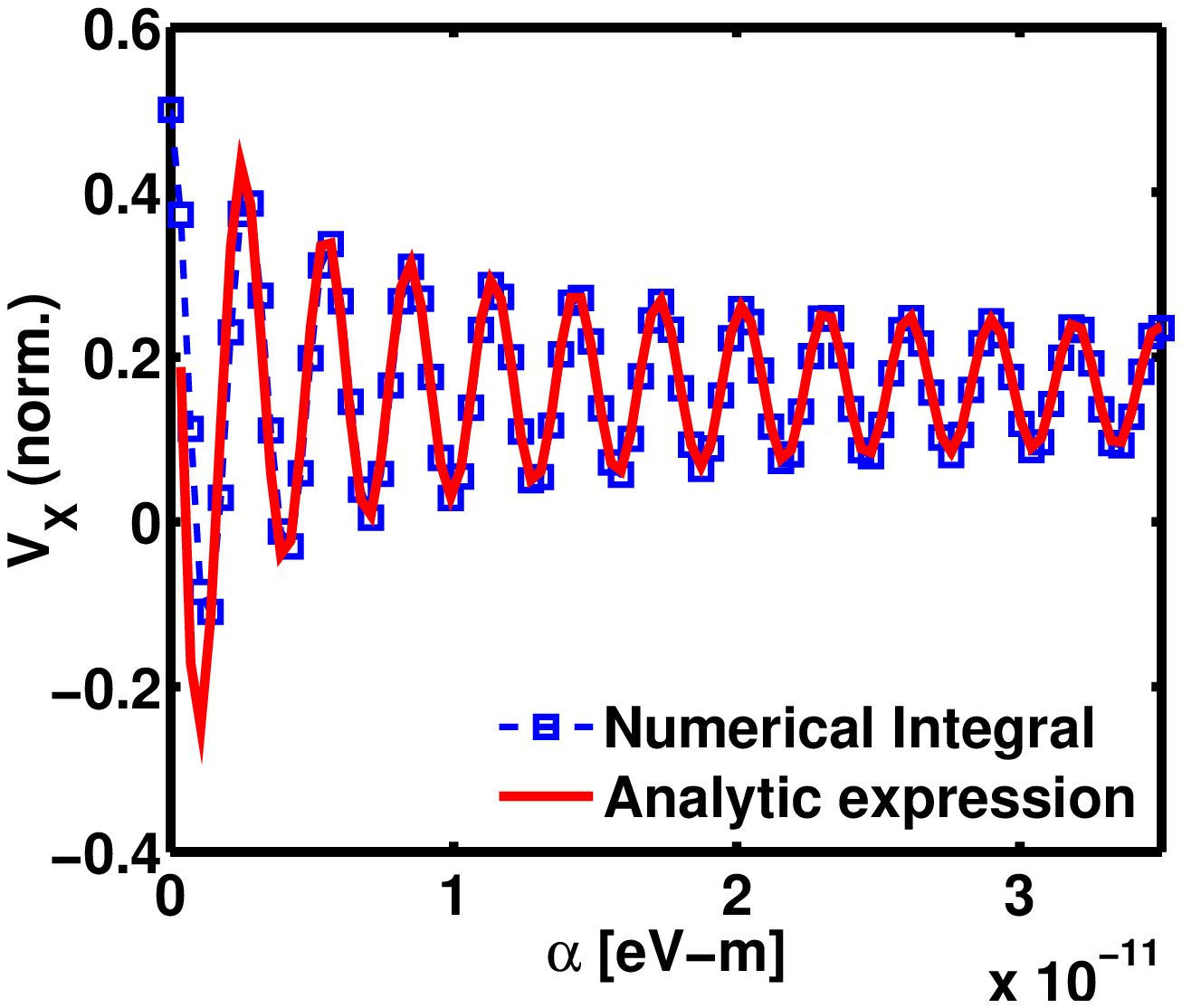}
\includegraphics[width=0.45\textwidth]{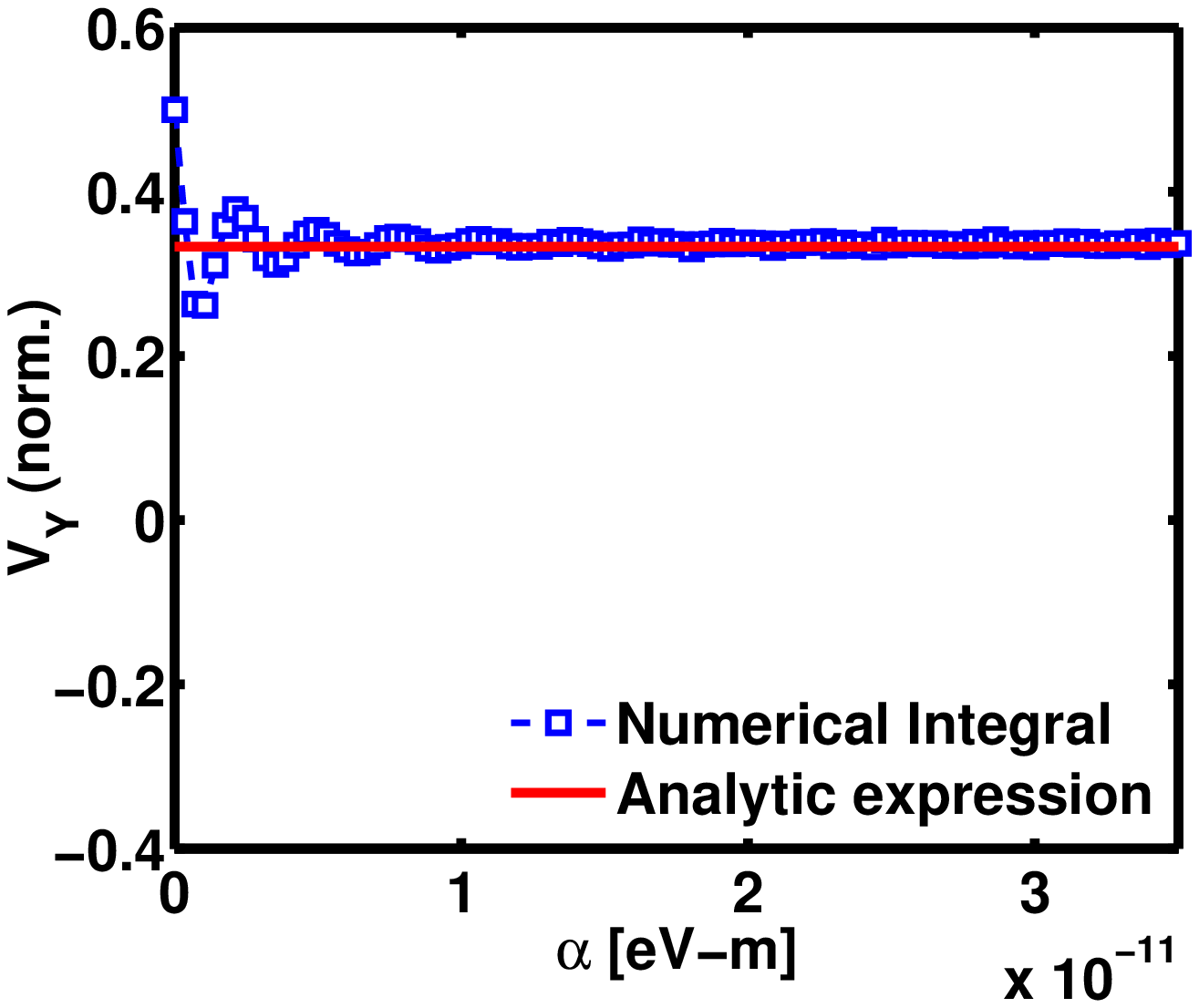}
\end{center}
\caption{{\it{Left}}: Numerical calculation (squares) of Eqs. 6, 5a versus analytical expression (solid) in Eq. 7a as a function of $\alpha$. {\it{Right}}: Numerical calculation (squares) of Eqs. 6, 5b versus analytical expression (solid) in Eq. 7b as a function of $\alpha$.}\label{stat_point_ext}
\end{figure}



From Eqs. 6 and 5a,
\begin{eqnarray}
V_{X} &=& \frac{1}{\pi}\int_{0}^{1}\mathrm{d}s\, V_{X0}(s) \nonumber \\
&=& \frac{B}{\pi}\int^1_0\mathrm{d}s\, s^2 + \frac{B}{\pi}\mathrm{Re}\left\{\int_{0}^{1}\mathrm{d}s\, (1-s^2)\exp\left(\frac{i\theta_L}{\sqrt{1-s^2}}\right)\right\} \nonumber \label{vxtheory5S}
\end{eqnarray} Noting that the phase has a stationary point at $s=0$ \cite{stat_point}, we expand it in Taylor's series around $s=0$ to obtain
\begin{eqnarray}
V_{X} &\simeq& \frac{B}{3\pi}+\frac{B}{\pi}\mathrm{Re}\left[\int_{0}^{0+\epsilon}\mathrm{d}s\, (1-s^2)\exp\left\{i\theta_L\left(1+\frac{s^2}{2}\right)\right\}\right] \nonumber\\
&\simeq& \frac{B}{3\pi}+\frac{B}{\pi}\mathrm{Re}\left\{\exp(i\theta_L)\int_{0}^{\infty}\mathrm{d}s\, \exp\left(i\theta_L\frac{s^2}{2}\right)\right\} \nonumber\\
&=& \frac{B}{3\pi}+\frac{B}{\pi}\mathrm{Re}\left\{\exp(i\theta_L)\frac{\exp(i\frac{\pi}{4})}{\sqrt{2\theta_L}}\Gamma\left(\frac{1}{2}\right)\right\} \nonumber \\
&=& \frac{B}{3\pi}+\frac{B}{\sqrt{2\pi\theta_L}}\cos(\theta_L+\frac{\pi}{4}) \nonumber
\label{vxtheory8a}
\end{eqnarray} as stated in Eq. 7a.

Similarly from Eqs. 6 and 5b,
\begin{eqnarray}
V_Y &=& \frac{B}{\pi}\int^1_0\mathrm{d}s\, (1-s^2)+\frac{B}{\pi}\mathrm{Re}\left\{\int^1_0\mathrm{d}s\, s^2\exp\left(\frac{i\theta_L}{\sqrt{1-s^2}}\right)\right\} \nonumber \\
&\simeq& \frac{2B}{3\pi} \nonumber
\label{vytheory8b}
\end{eqnarray} as stated in Eq. 7b.

\clearpage
\end{widetext}

\begin{thebibliography}{100}
\bibitem{koo_st_09} H. C. Koo {\it{et al.}}, Science {\bf{325}}, 1515 (2009).
\bibitem{datta_das_90} S. Datta and B. Das, Appl. Phys. Lett. {\bf{56}}, 665, (1990).
\bibitem{rso_ref} Yu. A. Bychkov and E. I. Rashba, JETP Lett. {\bf{39}}, 78 (1984).
\bibitem{nitta_sdh_97}  J. Nitta, T Akazaki, H. Takayanagi and T. Enoki, Phys. Rev. Lett. {\bf{78}}, 1335 (1997).
\bibitem{spin_inj_ref} See for example (a) A. Fert, J.-M. George, H. Jaffres and R. Mattana, IEEE Tran. Electron Devices, {\bf{54}}, 921 (2007) and (b) G. Schmidt, J. Phys. D: Appl. Phys. {\bf{38}}, R107 (2005), and the references therein.
\bibitem{vanWees_nl_03} See for example (a) M. Johnson and R. H. Silsbee, Phys. Rev. B {\bf{37}}, 5312 (1988) and (b) F. J. Jedema, M. S. Nijboer, A. T. Filip and B. J. van Wees, Phys. Rev. B {\bf{67}}, 085319 (2003).
\bibitem{zulicke} M. G. Pala, M. Governale, J. K\"{o}nig, and U. Z\"{u}licke, Europhys. Lett. {\bf{65}}, 850 (2004).
\bibitem{wolf_sbandy_dsarma} See for example (a) S. A. Wolf {\it{et. al.}}, Science {\bf{294}}, 1488 (2001), (b) S. Bandyopadhyay, Phys. Rev. B {\bf{61}}, 13813 (2000) and (c) S. D. Sarma, J. Fabian, X. Hu and I. Zutic, Solid State Comm. {\bf{119}}, 207 (2001).
\bibitem{dso_rso_ref} See for example (a) J. Luo, H. Munekata, F.F. Fang and P.J. Stiles, Phys. Rev. B {\bf{41}}, 7685 (1990) and (b) T. Koga, J. Nitta, T Akazaki, and H. Takayanagi, Phys. Rev. Lett {\bf{89}}, 046801 (2003).
\bibitem{takahashi} S. Takahashi and S. Maekawa, Phys. Rev. B {\bf{67}}, 052409 (2003).
\bibitem{stat_point} N. Bleistein and R. Handelsman (1986), Asymptotic Expansions of Integrals, Dover, New York; Jon Mathews and L. Robert (1970). Mathematical methods of physics (2nd ed.), New York, [NY.]: W. A. Benjamin.
\bibitem{sbandy_com_09} S. Bandyopadhyay, arXiv:cond-matt/0911.0210 (2009).
\bibitem{hbc_ref} See for example (a) B. K. Nicoli$\grave{c}$ and S. Souma, Phys. Rev. B {\bf{71}}, 195328 (2005) and (b) J. S. Jeong and H. W. Lee, Phys. Rev. B, {\bf{74}}, 195311 (2006).
\bibitem{hanle_ref} See for example (a) M. Johnson and R. H. Silsbee, Phys. Rev. Lett. {\bf{55}} 1790 (1985), (b) F. J. Jedema, H. B. Heersche, A. T. Filip, J. J. A. Baselmans and B. J. van Wees, Nature {\bf{416}}, 713 (2002) and (c) Lou X. et. al., Phys. Rev. Lett. {\bf{96}}, 176603 (2006).
\bibitem{awschalom} J. D. Koralek, C. P. Weber, J. Orenstein, B. A. Bernevig, S.-C. Zhang, S. Mack. and D. D. Awschalom, Nature {\bf{458}}, 610 (2009).
\bibitem{datta3} S. Datta, Nanoelectronics: A unified view Chapter 1, vol. I, the Oxford Handbook on Nanoscience and Nanotechnology, eds. A.V. Narlikar and Y. Y. Fu (arXiv:cond-matt/0809.4460, 2009).
\bibitem{datta2} S. Datta, {\it{Quantum Transport: Atom to Transistor}} (Cambridge University Press, Cambridge, 2005).
\end{thebibliography}
\end{document}